# High Quality, Transferrable Graphene Grown on Single Crystal Cu(111) Thin Films on Basal-Plane Sapphire

Kongara M. Reddy, Andrew D. Gledhill,

Chun-Hu Chen, Julie M. Drexler, and Nitin P. Padture[*]

Center for Emergent Materials

Department of Materials Science and Engineering

The Ohio State University

Columbus, OH 43210, USA

**Abstract**

The current method of growing large-area graphene on polycrystalline Cu surfaces (foils or thin films) and its transfer to arbitrary substrates is technologically attractive. However, the quality of graphene can be improved significantly by growing it on single-crystal Cu surfaces. Here we show that high quality, large-area graphene can be grown on epitaxial single-crystal Cu(111) thin films on reusable basal-plane sapphire ($\alpha$-$Al_2O_3$(0001)) substrates for transfer to another substrate. While enabling graphene growth on Cu single-crystal surfaces, this method has the potential to avoid the high cost and extensive damage to graphene associated with sacrificing bulk single-crystal Cu during graphene transfer.

______________________

[*] Electronic mail. padture.1@osu.edu

An unusual combination of tunable properties possessed by graphene makes this remarkable 2-D crystal of carbon an attractive material for a wide range of applications [1-4]. While the initial methods for synthesizing graphene — mechanical cleavage [1,5] and Si volatilization of SiC [6,7] — have been highly successful in enabling graphene science, they have fundamental limitations in terms of the size of the graphene that can be produced and the types of substrates on which it can be supported for technological applications. While there have been numerous other methods for making graphene or few layers graphene (FLG) [8], such as reduction of graphene oxide [9], stamping [10], cutting of carbon nanotubes [11,12] to name a few, chemical vapor deposition (CVD) of large-area graphene on metal surfaces is emerging as an attractive method from a technological standpoint. The growth of graphene or FLG by decomposition of hydrocarbons and/or segregation of dissolved C at surfaces of metals such as Pt [13], Ni [14-16], Ru [17,18] and Ir [19] has been reported. However, Cu appears to be the most promising metal for CVD because the resulting graphene is primarily monolayer, it can be grown over large areas, and Cu is relatively inexpensive [20-22]. Furthermore, the polycrystalline Cu substrate (foils or thin films) can be dissolved and the graphene can be transferred onto arbitrary substrates of interest under ambient conditions [20-23]. However, electron mobilities in CVD-grown graphene that has been transferred to $SiO_2$ substrates is low compared to those in graphene made by other methods [20,24]. While this can be attributed to the degradation of the CVD graphene as a result of the transfer process, the importance of starting off with the highest quality as-grown CVD graphene on Cu is being recognized [24,25].

In that context, the understanding of CVD growth of graphene on polycrystalline Cu foils is starting to emerge. Li *et al.* [26] have used $^{13}C$ isotopic labeling to show that the decomposition of hydrocarbons on Cu surface is the primary mechanism of self-limited growth of monolayer graphene. Note that, unlike in the case of Ni, carbon has extremely low solid-solubility in Cu at growth temperatures (~1000 °C), which precludes the growth of additional graphene layers from the bottom [26]. More recently, it has been shown that graphene nucleates in the form of numerous four-lobed "polydomain" islands at defects on the Cu surface (imperfections, grain boundary edges), which eventually coalesce to form a continuous layer of "polycrystalline" graphene with numerous domain boundaries [27-29].

To study the effect of lack of Cu grain boundaries, Gao *et al.* [24] have studied CVD-grown graphene on bulk single crystal Cu(111), and they have indicated that a reduction in the density of surface imperfections, including grain boundary edges, could reduce the concentration of nucleation sites. This in turn could reduce the concentration of domain boundaries resulting in improved graphene properties [24]. Rasool *et al.* [25] have studied the effect on CVD-grown graphene of surface facets (steps, edges, vertices) within single grains of polycrystalline Cu, and they have concluded that these types of surface imperfections do not affect growth of perfect graphene.

These results indicate that the use of high-purity, single crystal Cu may afford the possibility of growing the highest quality CVD graphene. However, there are two important practical drawbacks in using bulk single crystal Cu substrates. First, the cost of using bulk single-crystal Cu is prohibitive, because it is necessary to sacrifice the Cu substrate for subsequent transfer of the graphene. Second, the prolonged exposure to harsh chemical needed to dissolve bulk single-crystal Cu substrates completely can damage the graphene extensively before it is transferred to other useful substrates. In order to overcome these drawbacks, here we show that high quality, large-area graphene can be grown on epitaxial single-crystal Cu(111) thin films on reusable basal-plane sapphire (α-Al$_2$O$_3$(0001)) substrates for transfer to another substrate.

Several α-Al$_2$O$_3$(0001) substrates (12×12 mm$^2$; 1 mm thickness; surface roughness <1 nm) (MTI Corp., Richmond, CA) were thoroughly cleaned, and epitaxial Cu films were deposited via thermal evaporation using a method described by Katz [30]. The substrates were maintained at 250 °C inside a conventional thermal evaporator (BOC Edwards Ltd., Crawley, UK), and pure Cu pellets (99.999%; Kurt Lesker Company, Clairton, PA) were resistively heated at a pressure of ~10$^{-6}$ Torr for ~20 min. The thickness of the Cu film was determined to be ~400 nm using scanning electron microscopy (SEM).

The resulting Cu thin films were also characterized using X-ray diffraction (XRD), performed using a high-resolution triple-axis diffractometer (Bruker AXS, Karlsruhe, Germany). A locked-coupled scan was used after calibration. Analysis of the diffraction pattern was performed using DIFFRACplus EVA software (Bruker AXS, Karlsruhe, Germany). Figure 1a is a XRD pattern showing only (111) and (222) reflections, confirming single-crystal Cu(111) nature of the thin film. Electron backscatter diffraction (EBSD) was also performed to confirm the single-crystal nature of the Cu films, using a SEM (Quanta 200, FEI Company, Hillsboro, OR) equipped with a EBSD detector and software package (EDAX/TSL, Mahwah, NJ). The sample was tilted at 70° relative to the electron beam, and the resulting pattern had an estimated 99.9% confidence interval agreement with Cu(111). A 200×200 μm$^2$ area was selected and a map was collected with data being collected every 2 μm. Each datum point was collected six times, and the signals were averaged before indexing using the software package. These data are plotted in Fig. 1b which are clustered near 111 on the pole figure, further confirming the single-crystal nature of the Cu(111) thin film.

It is important to note that typical Cu thin films grown on a α-Al$_2$O$_3$(0001) substrate are polycrystalline, and that epitaxial growth of Cu is highly sensitive to the evaporation conditions. We have found that epitaxial growth occurs in a narrow substrate temperature range of 240 to 300 °C, and that the maximum epitaxial film thickness is limited to ~2 μm. Katz [30] has shown the presence of twin relationship in Cu deposited on α-Al$_2$O$_3$(0001), with the following epitaxial relationship between Cu and α-Al$_2$O$_3$:

$$(111)_{Cu} \| (0001)_{\alpha-Al2O3}; [2\bar{1}\bar{1}]_{Cu} \| [2\bar{1}\bar{1}0]_{\alpha-Al2O3} \qquad (1)$$

Graphene was grown on the Cu(111) thin films using a variation of the CVD process described by Li *et al.* [20]. Briefly, the α-Al$_2$O$_3$(0001) substrates with as-deposited Cu(111) thin films were placed in a CVD reactor consisting of an controlled-atmosphere quartz-tube furnace (Lindberg/Blue M, Asheville, NC). The tube was evacuated to a base pressure of 20 mTorr, and backfilled with H$_2$ to maintain a pressure of 250 mTorr. The samples were then heated to 1000 °C in 30 min and annealed for 20 min under a 10 sccm H$_2$ flow, followed by simultaneous introduction of high-purity CH$_4$ gas (35 sccm), achieving a final pressure of 700 mTorr. Graphene growth was carried out for 25 minutes at 1000 °C before allowing the furnace to cool to room temperature under flowing H$_2$+CH$_4$. It has been shown that the minimum thickness required to preclude dewetting of polycrystalline Cu films at graphene-growth temperature of ~1000 °C is ~500 nm [21,23]. It is possible that the epitaxial nature of the Cu(111) thin film resists dewetting, allowing the use thinner films (~400 nm) here.

The graphene grown on Cu(111) thin films was characterized by Raman spectroscopy (InVia Raman Microscope, Renishaw, Gloucestershire, UK) using a 514 nm wavelength laser (~2 μm spot size) at 1 mW power. Spectra from numerous locations on the graphene surface were collected, averaging 10 scans per location. Cu background was subtracted using the Wire 3.2 software (Renishaw, Gloucestershire, UK).  Figure 2a plots representative Raman spectra from three separate locations several mm apart showing the presence of 2D and G peaks.  Both peaks in all three spectra could be fitted to single Lorentzian functions confirming that the peaks are symmetric, and that the approx. 2D:G height ratios are 2:1.  The D peak attributed to defects in graphene is noticeably absent, indicating the high quality of single-layer graphene produced over a large area.  This was further confirmed by constructing a 2D-peak Raman map over a 70×70 µm$^2$ area.  Spectra in the Raman shift range 2600 to 2800 cm$^{-1}$ were collected at 5 µm intervals and the full-width-half-maxima (FWHM) of the 2D peaks were measured from Lorentzian fits.  The FWHM were then spatially mapped in Fig. 2b showing the tight distribution of 2D-peak FWHM (<45 cm$^{-1}$) over the entire area.

For comparison, CVD graphene was grown on a high-purity polycrystalline Cu foil of 25 µm thickness (Alfa Aesar, Ward Hill, MA) under identical conditions as above.  Figure 3a is an optical micrograph of that sample, showing a triple junction of grain-boundary edges in the underlying Cu foil.  Figure 3b is a Raman spectrum collected from the area indicated by the red circle in Fig. 3a.  In addition to the single-Lorentzian 2D and the G peaks in approx. 2:1 ratio, a prominent defect D peak is seen in Fig. 3b.  These results clearly demonstrate that the use of single crystal Cu thin films can result in high quality graphene over large areas, relative to graphene grown on a polycrystalline Cu foil under identical conditions.

Preliminary experiments indicate that the graphene grown on single crystal Cu thin films on sapphire can be readily transferred to other substrates using a variation of a method described by Levendorf *et al.* [21].

In summary, the use of epitaxial Cu(111) thin films on α-Al$_2$O$_3$(0001) enables CVD growth of high-quality graphene on single-crystal Cu surfaces over large areas.  At the same time, the thin film nature of the sacrificial Cu will allow its low-cost and easy removal for subsequent transfer of the graphene to other useful substrates.

Funding for this work was provided by the National Science foundation through the MRSEC (Grant No. 0820414) and the ENG-ECS (Grant No. 0925529) programs.

**Figure Captions**

Figure 1.  (a) XRD pattern of epitaxial Cu(111) on $\alpha$-Al$_2$O$_3$(0001) substrate indicating only 111 and 222 reflections. (b) EBSD data from epitaxial Cu(111) on $\alpha$ Al$_2$O$_3$(0001) substrate plotted on a pole figure showing only 111 orientation of Cu.

Figure 2.  (a) Raman spectra from three different regions of CVD graphene grown on epitaxial Cu(111) on $\alpha$-Al$_2$O$_3$(0001) substrate. The 2D and G peaks are marked, and the dashed line marks the position of the defect D peak which is absent. (b) Raman 2D-peak FWHM map showing a tight distribution (<45 cm$^{-1}$) over a 70×70 µm$^2$ area.

Figure 3.  (a) Optical micrograph (top view) of graphene grown on a polycrystalline Cu foil showing a triple junction of grain-boundary edges in the underlying Cu foil. (b) Raman spectra from the area indicated by the red circle in (b) showing the 2D and G peaks, and the defect D peak.

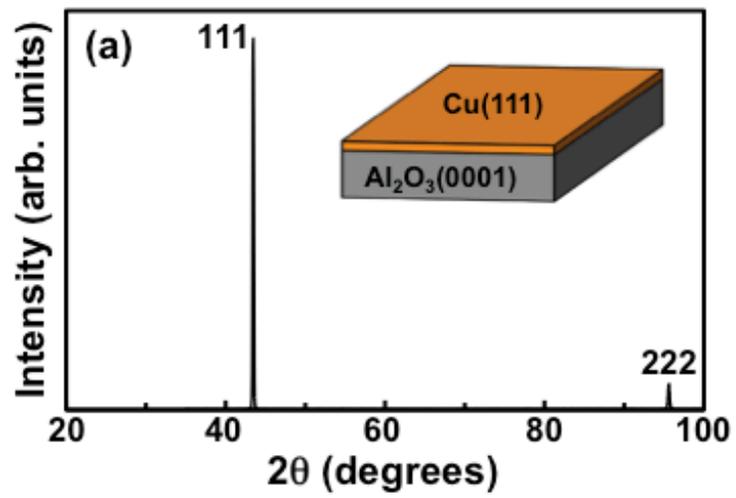

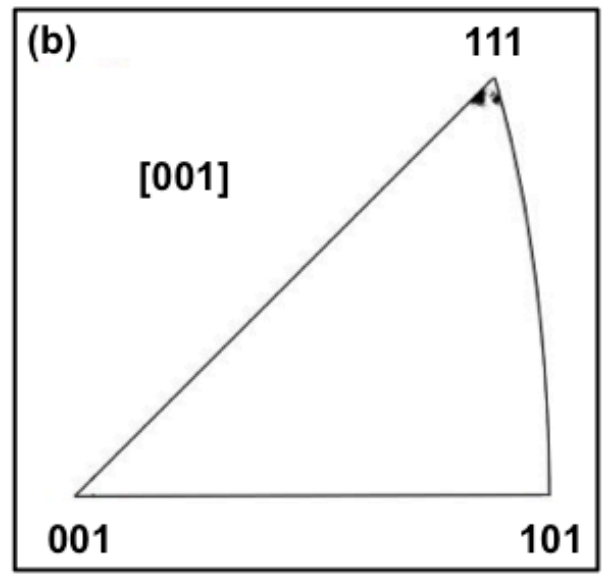

Reddy *et al.*
Figure 1

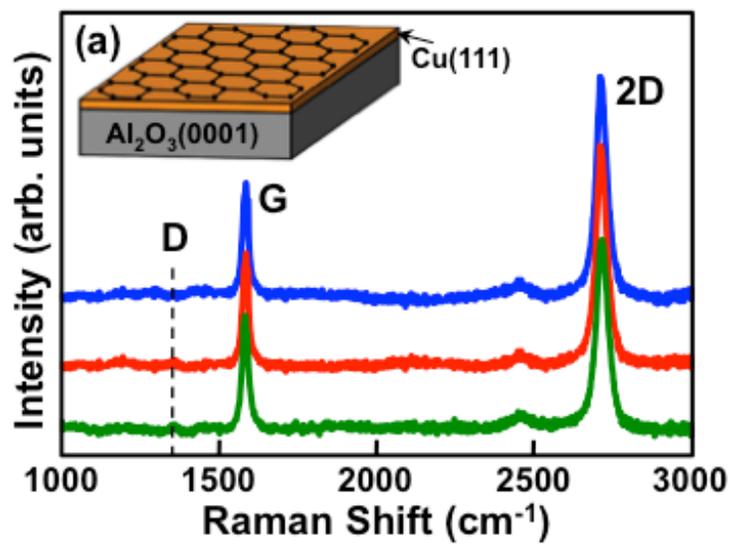
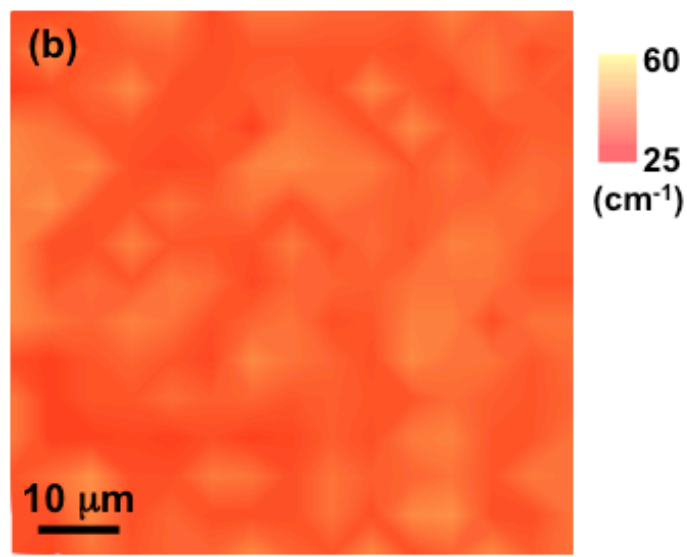

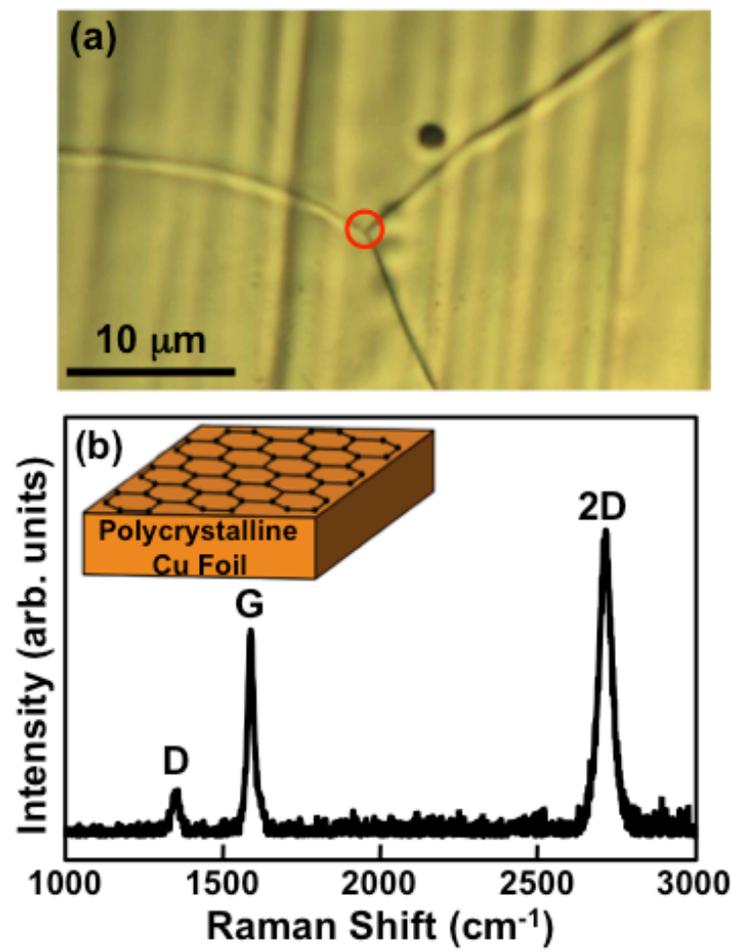
Reddy *et al.*
Figure 3